\documentclass[traditabstract]{aa}  

\usepackage{psfig}

\begin{document}

\title{Optical identification of X--ray source 1RXS J180431.1$-$273932
as a magnetic cataclysmic variable\thanks{Partly based on observations 
collected at the Italian Telescopio Nazionale Galileo, located at the 
Observatorio del Roque de los Muchachos (Canary Islands, Spain)}}


\author{N. Masetti\inst{1}, A.A. Nucita\inst{2,3} and P. Parisi\inst{1,4}}

\institute{
INAF -- Istituto di Astrofisica Spaziale e Fisica Cosmica di Bologna, via 
Gobetti 101, I-40129 Bologna, Italy
\and
Dipartimento di Matematica e Fisica ``Ennio De Giorgi", Universit\`a del
Salento, via per Arnesano, I-73100 Lecce, Italy
\and
INFN -- Istituto Nazionale di Fisica Nucleare, Sezione di Lecce, via per 
Arnesano, I-73100 Lecce, Italy
\and
INAF -- Istituto di Astrofisica e Planetologia Spaziali, via Fosso del 
Cavaliere 100, I-00133 Roma, Italy}

\offprints{N. Masetti (\texttt{masetti@iasfbo.inaf.it)}}
\date{Received 3 April 2012; accepted 5 July 2012}

\abstract{The X--ray source 1RXS J180431.1$-$273932 has been proposed 
as a new member of the symbiotic X--ray binary (SyXB) class of systems, 
which are composed of a late-type giant that loses matter to an extremely 
compact object, most likely a neutron star. In this paper, we present an 
optical campaign of imaging plus spectroscopy on selected candidate 
counterparts of this object. We also reanalyzed the available archival 
X--ray data collected with {\it XMM-Newton}. We find that the brightest 
optical source inside the 90\% X--ray positional error circle is 
spectroscopically identified as a magnetic cataclysmic variable (CV), most 
likely of intermediate polar type, through the detection of prominent 
Balmer, He {\sc i}, He {\sc ii}, and Bowen blend emissions. On either 
spectroscopic or statistical grounds, we discard as counterparts of 
the X--ray source the other optical objects in the {\it XMM-Newton} error 
circle. A red giant star of spectral type M5\,III is found lying just 
outside the X--ray position: we consider this latter object as 
a fore-/background one and likewise rule it out as a counterpart 
of 1RXS J180431.1$-$273932. The description of the X--ray spectrum of the 
source using a bremsstrahlung plus black-body model gives temperatures of
$kT_{\rm br} \sim$ 40 keV and $kT_{\rm bb} \sim$ 0.1 keV for these two 
components. We estimate a distance of $d \sim$ 450 pc and a 0.2--10 keV 
X--ray luminosity of L$_{\rm X} \sim$ 1.7$\times$10$^{32}$ erg s$^{-1}$ 
for this system and, using the information obtained from the X--ray 
spectral analysis, a mass $M_{\rm WD} \sim$ 0.8 $M_\odot$ for the 
accreting white dwarf (WD). We also confirm an X--ray periodicity of 494 s 
for this source, which we interpret as the spin period of the WD.
In summary, 1RXS J180431.1$-$273932 is identified as a magnetic CV
and its SyXB nature is excluded.}

\keywords{X--rays: individual: 1RXS J180431.1$-$273932 --- novae,
cataclysmic variables --- Stars: dwarf novae --- Techniques: 
spectroscopic --- Astrometry}

\titlerunning{The CV nature of 1RXS J180431.1$-$273932}
\authorrunning{N. Masetti et al.}

\maketitle

\section{Introduction}

Symbiotic X--ray binaries (SyXBs; see e.g. Masetti et al. 2006a) form a 
minor class of low mass X--ray binaries in which the compact accretor, 
most likely a neutron star (NS), receives matter from a red giant rather 
than from a late-type companion star on the main sequence (or possibly 
slightly evolved) and with mass of generally $\la$1 $M_\odot$. These 
objects are defined SyXBs by analogy with symbiotic binary systems, which 
are formed by an evolved late-type star and a white dwarf (WD).

There are currently only seven confirmed objects of this type known in 
the Galaxy: six cases listed in Masetti et al. (2007), Nespoli et al. 
(2010), and references therein, to which a newly-identified one, XTE 
J1743$-$363, was recently added (Smith et al. 2012).
It is therefore equally important to investigate possible new candidates
(cf. Masetti et al. 2011) and to spectroscopically confirm the
known candidates. In addressing the latter issue, Masetti
et al. (2012) found by using optical spectroscopy that the SyXB candidate 
2XMM J174016.0$-$290337 (also known as AX J1740.2$-$2903) proposed by 
Farrell et al. (2010) is actually a cataclysmic variable (CV) of dwarf 
nova type. Likewise, the availability of (sub)arcsec-sized X--ray positions
allows one to determine possible optical counterpart misidentifications,
especially in extremely crowded fields. This occurred in the 
case of IGR J16393$-$4643, for which a {\it Chandra} snapshot (Bodaghee et 
al. 2012) pinpointed the correct near-infrared counterpart and dismissed 
the one proposed by Nespoli et al. (2010) as a SyXB.

With the aim of confirming (or disproving) the nature of yet another 
SyXB candidate, we performed an optical imaging and spectroscopic campaign 
on two possible counterparts of the X--ray source 1RXS J180431.1$-$273932 
(Nucita et al. 2007); we also took this opportunity to reanalyze the 
X--ray data presented by those authors.

The X--ray object 1RXS J180431.1$-$273932, first detected in the {\it 
ROSAT} bright source survey (Voges et al. 1999), was subsequently observed 
on October 2005 with {\it XMM-Newton}. The main results of this 
observation, reported by Nucita et al. (2007), are: (i) the detection of 
an X--ray period of 494 s, most likely due to the spin of the compact 
accretor; (ii) the description of its X--ray spectrum in the 0.2-7 keV 
range in the form of a power-law with index $\Gamma \sim$ 1 plus a 
Gaussian emission line at $\sim$6.6 keV; and (iii) the detection, with the 
Optical Monitor (OM) onboard {\it XMM-Newton}, of an object with magnitude 
$v \sim$ 17.2 at a position consistent with the $\sim$2$''$-radius 
(1$\sigma$, corresponding to 3$\farcs$3 at the 90\% confidence level) 
X--ray error circle of the source.

Concerning the last point, Nucita et al. (2007) found that the OGLE 
catalog (Wray et al. 2004) reports a red optical object at $\sim$5$''$ 
from the X--ray position of the source that has a periodicity of about 
20.5 days in its $I$-band light curve. On the basis of the optical and 
near-infrared magnitudes of this object (assuming that the OM and OGLE 
sources are one and the same), Nucita et al. (2007) concluded that its 
colors are compatible with those of a red giant star of type M6\,III, thus 
making 1RXS J180431.1$-$273932 a viable SyXB candidate.

However, the non-negligible (albeit small) difference in the positions of 
the X--ray and the OGLE objects, together with the lack of optical 
spectroscopy for the latter, calls for an in-depth investigation of the 
properties of this source in the optical bands. In addition, the location 
of the source towards the Galactic center ($l$ = 3$\fdg$2; $b$ = 
$-$2$\fdg$9) suggests that the field crowdedness might produce source 
confusion when the positional uncertainty of an object is as large as a 
few arcsec. We therefore started an optical imaging and spectroscopic 
campaign to clarify the nature of 1RXS J180431.1$-$273932 using the 
Italian Telescopio Nazionale Galileo. We also decided to reanalyze here 
the {\it XMM-Newton} data first reported in Nucita et al. (2007) using 
updated software and response matrices and a more physical model to 
describe the X--ray spectrum.

The outline of the present paper is as follows: in Sect. 2, we describe 
our optical and X--ray observations, while Sect. 3 reports our results 
and Sect. 4 discusses them. Finally, in Sect. 5 we present our conclusions 
for this source.
 
\begin{figure}[!t]
\vspace{-3.6cm}
\hspace{-3.1cm}
\psfig{figure=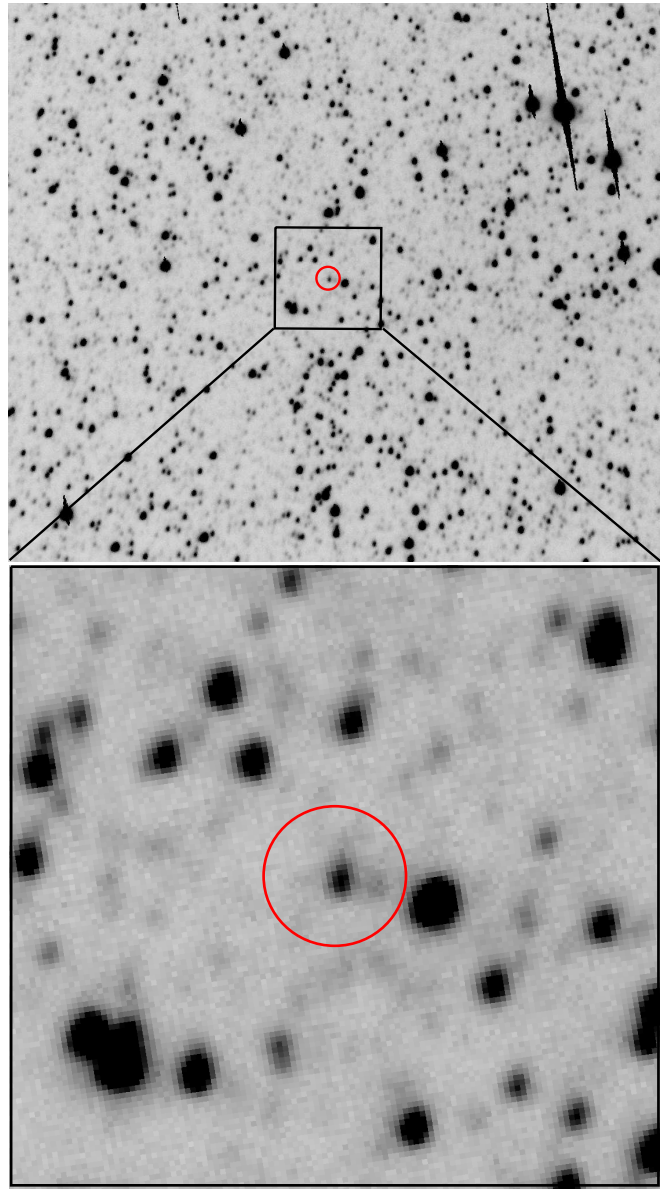,width=15cm,angle=0}
\vspace{-2cm}
\caption[]{{\it Upper panel:} TNG+DOLORES white (open) filter
image of the field of 1RXS J180431.1$-$273932 with a superimposed
3$\farcs$3-radius 90\% confidence level {\it XMM-Newton}
X--ray error circle. The field size is about 3$'$$\times$3$'$.
{\it Lower panel}: zoomed image of a 30$''$$\times$30$''$
box centered on the {\it XMM-Newton} position.
The actual counterpart (see text) is the object readily recognizable
within the circle; the red giant star mentioned by Nucita et al.
(2007) is the bright source just outside the circle on the right.
In both panels, north is at top and east is to the left.}
\end{figure}

\section{Observations}

\subsection{Optical}

All optical data presented here were acquired with the 3.58m Telescopio 
Nazionale Galileo (TNG), which is located in La Palma (Canary Islands, 
Spain) and equipped with the imaging spectrograph DOLORES.
This instrument carries a 2048$\times$2048 pixel back-illuminated, 
thinned E2V 4240 CCD.

\subsubsection{Imaging and astrometry}

To acquire a deeper and higher resolution image of the field of
1RXS J180431.1$-$273932 with respect to the one available from the DSS-II-Red
survey\footnote{Available at {\tt http://archive.eso.org/dss/dss}}, on 8 
September 2010 we obtained a white-filter snapshot of duration 10 s and 
start time 20:12:55 UT. In imaging mode, DOLORES can secure a field 
of 8$\farcm$6$\times$8$\farcm$6 with a scale of 0$\farcs$252 pix$^{-1}$.

The image thus acquired was then processed to obtain an astrometric 
solution based on 30 USNO-A2.0\footnote{The USNO-A2.0 catalog is 
available at \tt {http://archive.eso.org/skycat/servers/usnoa}} reference 
stars in the field of 1RXS J180431.1$-$273932. The conservative error in 
the optical position is 0$\farcs$252, which was added in quadrature 
to the systematic error in the USNO catalog (0$\farcs$25 according to 
Assafin et al. 2001 and Deutsch 1999). The final 1$\sigma$ uncertainty in 
the astrometric solution of the image is thus 0$\farcs$35.

Once we had determined the astrometry of the image, we superimposed it 
with the X--ray error circle determined with the {\it XMM-Newton} data 
presented in Nucita et al. (2007; see also Sect. 3). This clearly 
encircles one relatively bright object (see Fig. 1); a brighter source is 
also present just outside the {\it XMM-Newton} error circle, west of it.

To get an estimate of the image depth, we also performed a photometric 
study of the image itself. Owing to the field crowdedness (see Fig. 1), we 
chose standard point spread function (PSF) fitting technique by using the 
PSF-fitting algorithm of the DAOPHOT II image data-analysis package 
(Stetson 1987) running within MIDAS\footnote{MIDAS (Munich Image Data 
Analysis System) is developed, distributed and maintained by the European 
Southern Observatory and is available at {\tt 
http://www.eso.org/sci/software/esomidas/}}.

No absolute calibration in magnitude can be given since this is a
white filter frame. However, a 3$\sigma$ limit of $\sim$3 magnitudes
fainter than that of the brightest object inside the {\it XMM-Newton}
error circle can be attributed to the depth of the image.

To clarify the nature of the two objects mentioned above, and to 
eventually determine which of the two (if any) is the actual counterpart 
of the X--ray source, we decided to undertake optical spectroscopy of 
both.

Additionally, one may note the presence of fainter objects inside the 
X-ray error circle (zoom-in of Fig. 1): we later discuss them and exclude 
their connection with 1RXS J180431.1$-$273932 in Sect. 3.1.

\subsubsection{Spectroscopy}

Optical spectroscopic data of the two brightest optical sources 
mentioned in the previous subsection were acquired on 21 August 2011 using 
the LR-B grism and a 1$\farcs$5 slit: this setup provided a dispersion of 
2.7 \AA/pixel and a nominal wavelength coverage between 3700 \AA~and 8100 
\AA. The total exposure time was 3$\times$20 min centered on 21:24 UT. The 
spectrograph slit was suitably oriented to acquire the spectra of both 
objects simultaneously.

The spectra, after correction for flat-fielding, bias, and cosmic-ray 
rejection, were background-subtracted and optimally extracted (Horne 1986) 
using IRAF\footnote{IRAF is the Image Analysis and Reduction Facility made 
available to the astronomical community by the National Optical Astronomy 
Observatories, which are operated by AURA, Inc., under contract with the 
U.S. National Science Foundation. It is available at {\tt 
http://iraf.noao.edu/}}. Wavelength calibration was performed using 
comparison-lamp exposures acquired soon after each on-target spectroscopic 
exposure, while flux calibration was accomplished by observing the 
spectroscopic standard-star Feige 110 (Hamuy et al. 1992, 1994).

The wavelength calibration uncertainty was $\sim$0.5 \AA; this was checked 
by using the positions of background night-sky lines. Spectra from the 
same object were then stacked together to increase the final 
signal-to-noise ratio.

\subsection{X--rays}

As mentioned above, we took this opportunity to reanalyze the X--ray data 
collected by the {\it XMM-Newton} satellite towards the source 1RXS 
J180431.1$-$273932. The source had been observed for $\simeq 100$ ks 
(Observation ID 30597) with both the EPIC MOS and pn cameras 
(Str\"uder et al. 2001; Turner et al. 2001) in thin filter mode. We 
processed the observation data files (ODFs) using the {\it XMM-Newton} 
Science Analysis System (SAS\footnote{{\tt http://xmm.esa.int/sas/}} 
version 11.0.0), together with the latest calibration constituent files. 
After processing the raw data via the standard {\it emchain} and {\it 
epchain} tasks, we were left with adequate event lists to use in 
the subsequent spectral and timing analysis. The J2000 X--ray coordinates 
of 1RXS J180431.1$-$273932 were determined once again by using the {\it 
edetect\_chain} tool in the MOS 1 and MOS 2 images in the 0.3--8.0 
keV band. The coordinates in output for the two cameras were then averaged 
in order to get the best estimate of the target position.

\subsubsection{X--ray spectral analysis}

For the spectral analysis, we further screened the event files by 
rejecting time intervals affected by high levels of background. These 
intervals (more evident in the energy range 10--12 keV) were flagged, 
strictly following the recipe described in the XRPS user's 
manual\footnote{Available at:\\ {\tt 
http://xmm.esac.esa.int/external/xmm\_user\_support/ \\
/documentation/rpsman/index.html}}, i.e. by selecting a threshold of 
0.4 counts s$^{-1}$ and 0.35 counts s$^{-1}$ for the the pn and MOS 
cameras, respectively. 

After inspecting by eye that the screening procedure effectively removed
the periods of high background activity, the good time intervals resulted 
in effective exposures of $\simeq$96 ks, $\simeq$98 ks, and $\simeq$94 ks 
for the MOS 1, MOS 2, and pn cameras, respectively.

The source spectra (one for each EPIC camera) were extracted from a 
circular region centered on the target position, while the background was 
extracted from source-free circular regions on the same chip and, 
where possible, at the same vertical location of the source extraction 
regions. Both the source and background extraction regions had a radius of 
$64''$. Particular attention had to be paid in the case of the pn 
camera since the target source is localized on a chip gap. In this case, 
the background extraction region was chosen on one of the CCDs in a 
position close to the target and free from other X--ray sources. For 
the pn data, we decided to accept only single events\footnote{This 
was chosen because the energy calibration for single events is slightly 
better than that for double ones, see e.g. the XRPS user's manual.} 
({\tt PATTERN = 0}), while, in the case of the two MOS, all valid patterns 
({\tt PATTERN $\leq$ 12}) were included. In all cases, we also added the 
selection {\tt FLAG = 0}.

\subsubsection{X--ray timing analysis}

The timing analysis was performed without applying any selection for good 
time intervals in order to avoid gaps that could introduce spurious 
effects. The synchronized source and background light curves were 
extracted in the 0.3--8 keV energy band for the three EPIC cameras. The 
light curves were additionally corrected (for absolute and relative 
corrections, see the XRPS user's manual) by using the {\it epiclccorr} 
task and then averaged in order to increase the signal-to-noise ratio. We 
then searched for any periodicity of between 5 s and 10 h by using the 
Lomb-Scargle method (Lomb 1976; Scargle 1982).

\section{Results}

\subsection{Optical}

\begin{figure*}
\hspace{-.5cm}
\mbox{\psfig{file=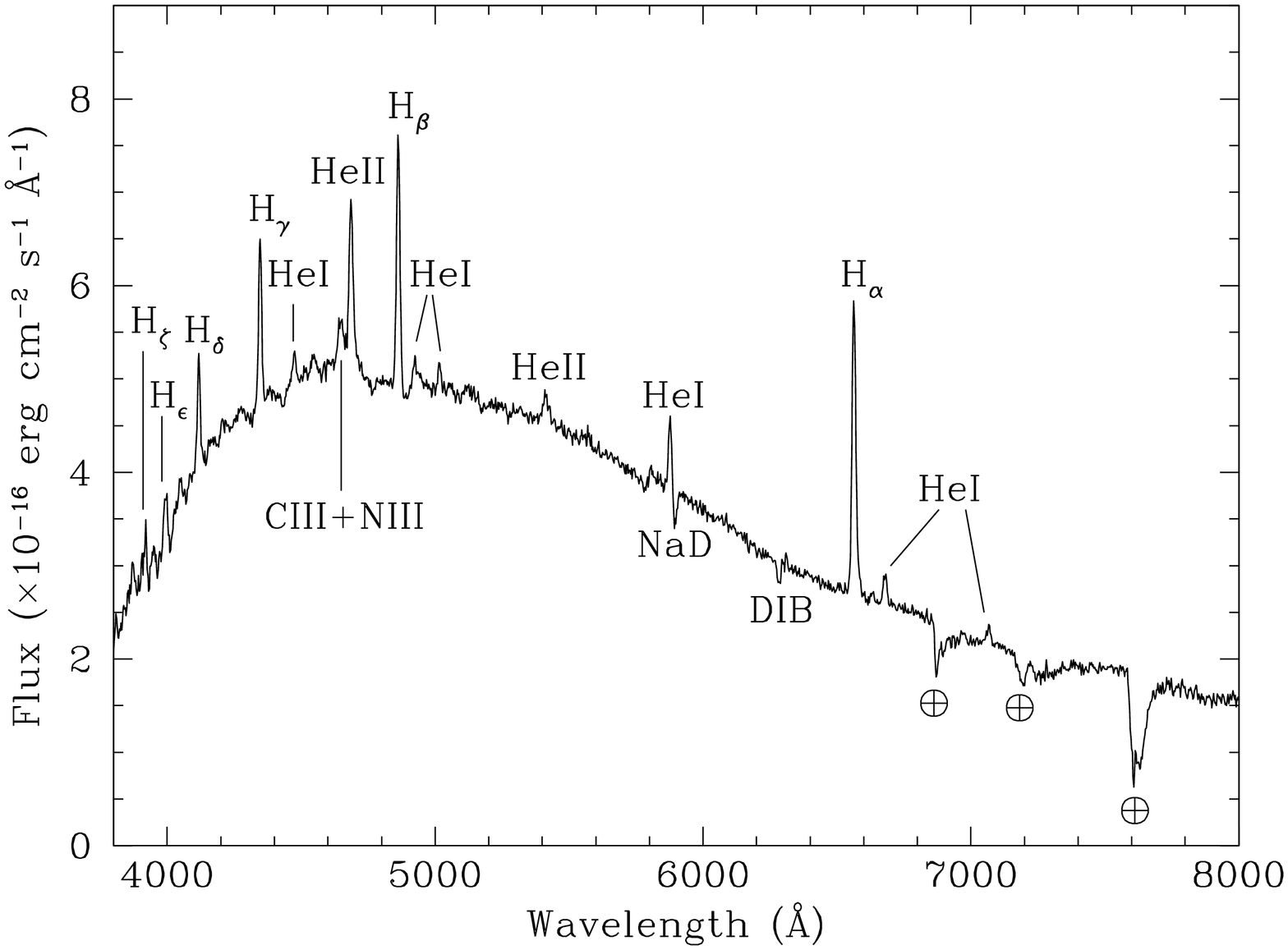,width=9.5cm,angle=270}}
\mbox{\psfig{file=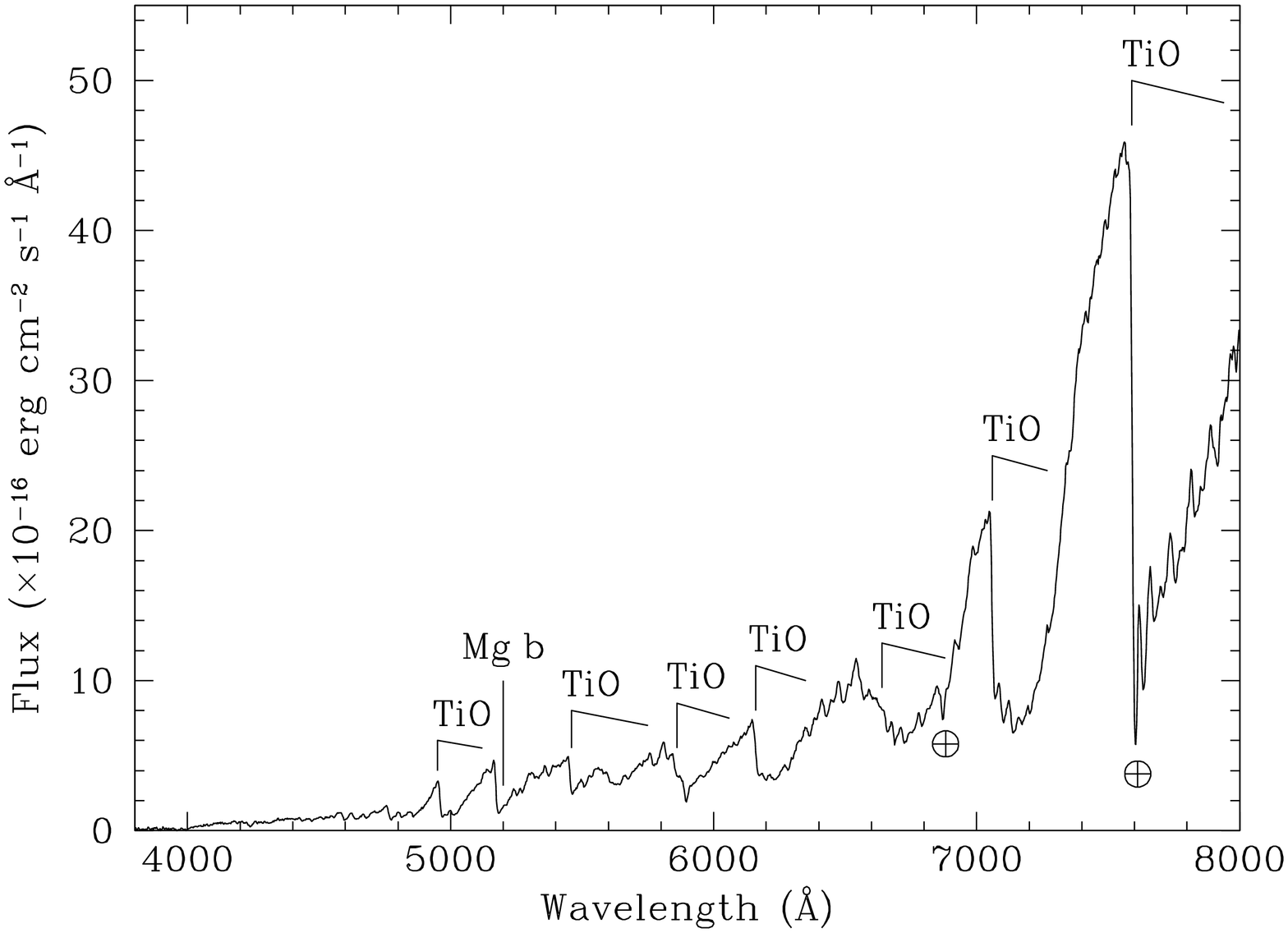,width=9.5cm,angle=270}}
\vspace{-.5cm}
\caption[]{The 3800--8000 \AA~optical spectra of the two main optical 
sources mentioned in this paper as possible optical counterparts of the 
X--ray source 1RXS J180431.1$-$273932 (see Fig. 1), that is, the 
most evident of the objects inside the X--ray error circle (left 
panel) and the bright one just outside it to the west (right 
panel). The former has a spectrum typical of a magnetic CV, while the 
latter shows the characteristics of a late-type giant star (see text). In 
both spectra, the telluric absorption bands are marked with the symbol 
$\oplus$.}
\end{figure*}

As mentioned in the previous section, our optical imaging shows that a
relatively bright object is found within the {\it XMM-Newton} error 
box. This source has coordinates RA = 18$^{\rm h}$ 04$^{\rm m}$ 30$\fs$44, 
Dec = $-$27$^{\circ}$ 39$'$ 32$\farcs$1 (J2000). It lies 0$\farcs$9 arcsec 
from the center of the X--ray error circle (Nucita et al. 2007; see also 
Sect. 3.2), thus well within it.
An additional, brighter source is present slightly outside the X--ray 
error circle and has coordinates consistent with the OGLE one reported in 
Nucita et al. (2007). This object is also reported in the 2MASS catalog 
(Skrutskie et al. 2006) as 2MASS J18043013$-$2739340.

The optical spectra of these two objects are reported in Fig. 2. The 
source inside the X--ray error circle (left panel) shows a number of 
emission lines, among which we identify the Balmer ones (up to at least 
H$_\zeta$), He {\sc i}, He {\sc ii}, and the Bowen blend around 4640 \AA; 
all features lie at $z$ = 0, indicating that this is a Galactic object. 
Fluxes and EWs of the main emission lines of this object are reported in 
Table~1.

These spectral characteristics are typical of CVs of dwarf nova type (see 
e.g. Masetti et al. 2006b); moreover, the Balmer decrement clearly appears 
to be negative, the He{\sc ii}$\lambda$4686/H$\beta$ EW ratio is larger 
than 0.5, and the EWs of He {\sc ii} and H$\beta$ are around 10 \AA~(see 
Table~1); all this indicates that this source is quite possibly a magnetic 
CV belonging to the intermediate polar (IP) subclass (see Warner 1995 and 
references therein) and that is not strongly affected by interstellar 
reddening.

\begin{table*}
\caption[]{List of the main results concerning the CV discovered within
the {\it XMM-Newton} error circle of source 1RXS J180431.1$-$273932 (see 
Fig. 1).}
\scriptsize
\vspace{-.3cm}
\begin{center}
\begin{tabular}{cccccccccr}
\noalign{\smallskip}
\hline
\hline
\noalign{\smallskip}
\multicolumn{2}{c}{H$_\alpha$} & \multicolumn{2}{c}{H$_\beta$} & 
\multicolumn{2}{c}{He {\sc ii} $\lambda$4686} &
$V$ & $A_V$ & $d$ & \multicolumn{1}{c}{$L_{\rm X}$} \\
\cline{1-6}
\noalign{\smallskip}
 EW & Flux & EW & Flux & EW & Flux & mag & (mag) & (pc) & (0.2--10 keV) \\

\noalign{\smallskip}
\hline
\noalign{\smallskip}

 20.8$\pm$1.0 & 5.6$\pm$0.3 & 8.9$\pm$0.4 & 4.4$\pm$0.2 & 6.3$\pm$0.4 & 3.0$\pm$0.2 &
 $\sim$17.3 & $\sim$0 & $\sim$450 & 1.7 \\

\noalign{\smallskip}
\hline
\noalign{\smallskip}
\multicolumn{10}{l}{Note: equivalent widths are expressed in \AA~and line 
fluxes are in units of 10$^{-15}$ erg cm$^{-2}$ s$^{-1}$,}\\
\multicolumn{10}{l}{whereas the X--ray luminosity observed with {\it 
XMM-Newton} is in units of 10$^{32}$ erg s$^{-1}$.} \\
\noalign{\smallskip}
\hline
\hline
\noalign{\smallskip}
\end{tabular}
\end{center}
\end{table*}

The spectrum of source 2MASS J18043013$-$2739340 (Fig. 2, right panel) 
shows instead the typical features of red giants (namely, a series of TiO 
bands) and no apparent Balmer lines in either emission or absorption. 
Using the Bruzual-Persson-Gunn-Stryker (Gunn \& Stryker 1983) spectroscopy 
atlas, we find that the spectrum of the source is strikingly similar to 
that of star BD $-$02$\fdg$3886, of spectral type M5 III. This supports 
the preliminary classification proposed by Nucita et al. (2007), which was 
based on the optical and near-infrared colors of this object.

Unfortunately, the white filter image we acquired does not allow us to
determine any reliable optical magnitude for the two objects, owing to the 
breadth of the filter used and the lack of calibration stars.
We therefore used the spectral flux information to extract a $V$-band 
magnitude for both the CV and the red giant. Using the flux-to-magnitude 
conversion factors of Fukugita et al. (1995), we found that $V_{\rm CV} 
\sim$ 17.3 and $V_{\rm RG} \sim$ 17.5.

Although the systematic uncertainties tied to this procedure may 
admittedly be large (possibly up to 20\%), we were able to perform here a 
broad determination of the optical magnitude of these objects.

These values, assuming absolute magnitudes of M$_V \sim$ +9 for the CV 
(Warner 1995) and M$_V \sim$ +0.7 (Th\'e et al. 1990) for a red giant of 
spectral type M5\,III, give the distances to the two sources of 
$d_{\rm CV} \sim$ 450 pc and $d_{\rm RG} \sim$ 23 kpc. We stress that 
these values (especially the one for the red giant) should conservatively 
be considered as upper limits, as the effect of the unknown amount of 
interstellar absorption along the line of sight was not accounted for in 
any of the two cases.

As mentioned in the previous section, a few additional fainter
objects are present in the {\it XMM-Newton} X--ray error circle:
one lying along the connecting line between the CV and the red giant,
and one north of the CV (plus possibly a further, fainter one east
of the CV).
We can exclude any connection of these optical sources with 1RXS 
J180431.1$-$273932 based on the following considerations.

First, owing to its sky position, we serendipitously obtained the
spectrum of the source located between the CV and the red giant at
the same time as we acquired spectroscopy of these two objects.
This is reported in Fig. 3: the presence of the G band at 4304 \AA,
the Mg {\sc i} band at 5175 \AA, and the Na {\sc i} doublet at
5890 \AA, along with the absence of any peculiar spectral features,
allow us to classify this source as a star of G type.

One can note that the optical spectral emission peak of this object 
lies around 5900 \AA, which is more redward than expected in a star of 
this spectral type ($\sim$5000 \AA; Jaschek \& Jaschek 1987): this is most 
likely due to the blue light being absorbed by interstellar dust along the 
line of sight. Moreover, this source also shows very weak Ca H+K lines 
around 4000 \AA. Although this may appear somewhat unusual for a G-type 
star, `weak-line' objects of this kind with abundance anomalies in their 
chemical composition are known (see e.g. Jaschek \& Jaschek 1987); 
alternatively, these absorption lines may be filled up by emission lines 
connected with the star's chromospheric activity (Houdebine et al. 2009 
and references therein). None of these peculiarities, however, implies 
that the observed X--ray emission could be produced by this object.

\begin{figure}[th!]
\hspace{-.3cm}
\mbox{\psfig{file=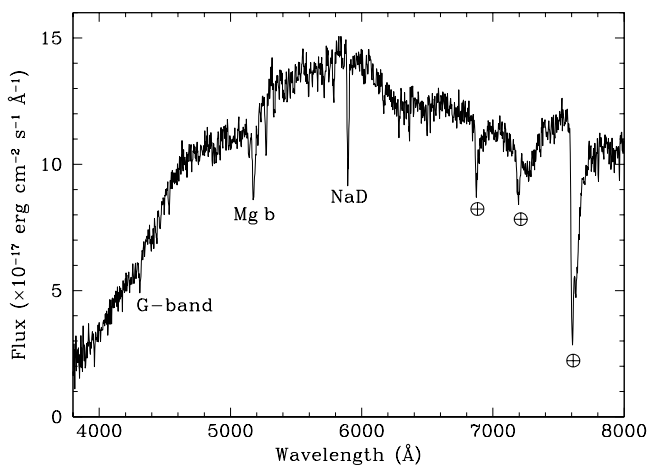,width=9.5cm,angle=270}}
\vspace{-.5cm}
\caption[]{The 3800--8000 \AA~optical spectrum of the faint object
located between the CV and the red giant as seen in the zoom-in of the
field of 1RXS J180431.1$-$273932 (Fig. 1, lower panel). The spectrum is
typical of a normal star of G type. Telluric absorption bands are marked 
with the symbol $\oplus$.}
\end{figure}

On the basis of statistical considerations, we next studied the 
association of the faintest object(s) both lying inside the X--ray error 
box of 1RXS J180431.1$-$273932 and of brightness $\sim$2.5 mag fainter 
than the CV in our white filter image. From the number density analysis of 
observed sources in the imaging frame, we expect to randomly find 1.2 
objects of this magnitude within an area of the size of the {\it 
XMM-Newton} X--ray error circle.

Approaching this issue from a different side, by considering the spatial 
density of CVs in the Galaxy (Rogel et al. 2008), we find that the 
probability of finding by chance a CV within 450 pc of the Earth in a sky 
area of size equal to the X--ray error box of this source is less than 
6$\times$10$^{-7}$.

All of the above allows us to say that 1RXS J180431.1$-$273932 can be 
identified as a magnetic CV beyond any reasonable doubt.

\subsection{X--rays}

\begin{table*}[t!]
\caption[]{X--ray spectral fits for 1RXS J180431.1$-$273932.}
\vspace{-.3cm}
\hspace{-.2cm}
\scriptsize
\begin{tabular}{lllllllllllll}
\noalign{\smallskip}
\hline
\hline
\noalign{\smallskip}
\multicolumn{13}{c}{Model in XSPEC: phabs*(power-law + gaussian)}\\
\hline
         &n$_{{\rm H}}$          & ${\rm \Gamma}$  & N$_{{\rm \Gamma}}$&E${_{\rm L}}$&${\rm \sigma_L}$&N${_{\rm L}}$&  $\chi^2_{\nu}$& d.o.f. &\multicolumn{4}{c}{ }  \\
         &                       &                 &($10^{-4}$)        &(keV)        &(keV)           &($10^{-4}$)  &                &       &\multicolumn{4}{c}{ }  \\
\hline
\multicolumn{13}{c}{}\\
$\overline{\rm{M}}$      &$0.15^{+0.01}_{-0.01}$&$1.05^{+0.01}_{-0.01}$&$4.0^{+0.1}_{-0.1}$&$6.57^{+0.01}_{-0.01}$&$0.23^{+0.05}_{-0.05}$&$0.17^{+0.04}_{-0.04}$&1.09&396& \multicolumn{4}{c}{ }  \\
\hline
\multicolumn{13}{c}{}\\
$\overline{\rm{M,p}}$  &$0.16^{+0.02}_{-0.02}$&$1.08^{+0.02}_{-0.02}$&$3.8^{+0.1}_{-0.1}$&$6.57^{+0.05}_{-0.05}$&$0.21^{+0.05}_{-0.05}$&$0.16^{+0.03}_{-0.03}$&2.00&575& \multicolumn{4}{c}{ }  \\
\hline
\multicolumn{13}{c}{}\\
$\overline{\rm{M}}$,p  &$0.15^{+0.01}_{-0.01}$&$1.07^{+0.02}_{-0.02}$&$4.0^{+0.1}_{-0.1}$&$6.58^{+0.05}_{-0.05}$&$0.23^{+0.04}_{-0.04}$&$0.18^{+0.06}_{-0.06}$ &1.08&573& \multicolumn{4}{c}{ }  \\
         &             &             &$3.0^{+0.1}_{-0.1}$&             &             &$0.17^{+0.05}_{-0.05}$ &    &   & \multicolumn{4}{c}{ }  \\
\hline
\multicolumn{13}{c}{}\\
\multicolumn{13}{c}{Model in XSPEC: phabs*pcfabs*(bremsstrahlung + black-body + gaussian)}\\
\hline
    & n$_{\rm H}$        & n$_{\rm H}$         &$CF$& $kT_{\rm br}$& N$_{\rm br}$  & $kT_{\rm bb}$ & N$_{{\rm bb}}$&E${_{\rm L}}$ & ${\rm \sigma_L}$&  N${_{\rm L}}$&  $\chi^2_{\nu}$& d.o.f.  \\
    & (Phabs)            & (Pcfabs)            &    & (keV)        & ($10^{-4}$)   & (keV)         &               &(keV)         & (keV)           &             &          &   \\
\hline
\multicolumn{13}{c}{}\\
$\overline{\rm{M}}$    &[0.23]&$6.0^{+2.4}_{-2.1}$&$0.28^{+0.07}_{-0.15}$&$40^{+50}_{-20}$ &$12.3^{+4.0}_{-0.7}$ &$0.12^{+0.01}_{-0.01}$            &$217^{+160}_{-100}$ &$6.59^{+0.07}_{-0.07}$&$0.27^{+0.05}_{-0.05}$&$0.22^{+0.05}_{-0.05}$      &1.05&393\\
\hline
\multicolumn{13}{c}{}\\
$\overline{\rm{M,p}}$&[0.23]&$5.5^{+2.0}_{-2.0}$&$0.32^{+0.08}_{-0.10}$&$30^{+12}_{-8}$ &$11.01^{+0.05}_{-0.03}$ &$0.13^{+0.02}_{-0.02}$         &$140^{+116}_{-70}$ &$6.59^{+0.05}_{-0.05}$&$0.26^{+0.05}_{-0.05}$&$0.22^{+0.04}_{-0.04}$         &1.93&572\\
\hline
\multicolumn{13}{c}{}\\
$\overline{\rm{M}}$,p&[0.23]&$5.6^{+1.5}_{-2.1}$&$0.37^{+0.06}_{-0.10}$&$40^{+60}_{-20}$&$12.0^{+1.6}_{-0.4}$    &$0.13^{+0.01}_{-0.01}$&$187^{+122}_{-84}$          &$6.60^{+0.05}_{-0.05}$&$0.27^{+0.05}_{-0.05}$&$0.23^{+0.05}_{-0.05}$         &1.03&569\\
                      &      &                   &                      &                &$9.0^{+0.7}_{-0.4}$     &                      &$140^{+120}_{-60}$          &                      &                      &$0.20^{+0.05}_{-0.05}$         &    &   \\
\noalign{\smallskip}
\hline
\noalign{\smallskip}
\multicolumn{13}{l}{Note: When the reduced $\chi^2_{\nu}$ is close to the value of 2.0, the errors (90\% 
confidence level) associated with the model parameters were obtained by using}\\
\multicolumn{13}{l}{the {\it steppar} command within XSPEC. Hydrogen column-densities are in units of 10$^{22}$ 
cm$^{-2}$. Fixed values in the fits are indicated in square brackets.} \\
\multicolumn{13}{l}{In the first column, the character M represents the MOS 1 and MOS 2 data sets, 
and p stands for the pn data. A line over the character (or set} \\
\multicolumn{13}{l}{of characters) indicates that, in the corresponding fit, 
the model normalizations were assumed to be identical for the overlined 
data sets.} \\ 
\noalign{\smallskip}
\hline
\hline
\noalign{\smallskip}
\end{tabular}
\end{table*}

From the averaged MOS 1 and MOS 2 data in the 0.3--8.0 keV band,
we found that the X--ray source lies at the position (J2000) 
RA = 18$^{\rm h}$ 04$^{\rm m}$ 30$\fs$48, Dec = $-$27$^\circ$ 39$'$ 
32$\farcs$76, which is thus in full agreement with Nucita et al. (2007).
The (1$\sigma$) statistical uncertainty in the source position, as
determined by the {\it edetect$\_$chain} task, is 0$\farcs$05. This value
is much smaller than the total 1$\sigma$ absolute astrometric accuracy of 
the MOS cameras, which was found to be 2$''$ (see e.g. Kirsch et al. 2004
and Guainazzi 2011\footnote{See also \\ {\tt 
http://xmm.esac.esa.int/external/
\\ xmm$\_$data$\_$analysis/sas$\_$workshops/sas$\_$ws11$\_$files/}}). We
therefore associate a 90$\%$ confidence level error of 3$\farcs$3 with 
both of the X--ray coordinates of 1RXS J180431.1$-$273932.

The spectra (binned with at least 25 counts per energy interval) were 
loaded into the fitting package XSPEC (Arnaud 1996), version 12.0.0.

The data were first fitted with a phenomenological model consisting of
an absorbed power-law to which a Gaussian line was added. This model was 
characterized by six free parameters (see Nucita et al. 2007 for 
further details), i.e. the hydrogen column-density n$_{\rm H}$, the photon 
index $\Gamma$, the line central energy E$_{\rm L}$, the emission-line 
width $\sigma_{\rm L}$, and the normalization of the Gaussian line and 
the power-law component N$_{\rm L}$ and N$_{\Gamma}$, respectively. 

On the basis of our results of Sect. 3.1, we then considered a more 
physical model related to the basic picture for the accretion onto a 
magnetized CV (e.g. Mouchet et al. 2008), i.e. absorbed bremsstrahlung and 
black-body components with a Gaussian line accounting for the emission 
feature observed at $\simeq$6.6 keV. A neutral absorber partially covering 
the source was also used to describe the intrinsic absorption, as 
sometimes seen in the spectra of similar objects (Rana et al. 2005; Homer 
et al. 2006).

Practically, the physical model was made of nine free parameters, 
i.e. the intrinsic hydrogen column-density n$_{\rm H}$, the covering 
factor $CF$, the temperatures of the bremsstrahlung and black-body 
components ($kT_{\rm br}$ and $kT_{\rm bb}$), together with their 
corresponding normalizations (N$_{\rm br}$ and N$_{\rm bb}$), the 
Gaussian line central energy E$_{\rm L}$, its width $\sigma_{\rm L}$, and 
the associated normalization N$_{\rm L}$. In the physical model, we noted 
that a stable fit was reached by fixing the neutral hydrogen 
column-density\footnote{For comparison, the neutral hydrogen 
column-density in the direction of the target as provided by the 
``N$_{\rm H}$" online calculator ({\tt 
http://heasarc.nasa.gov/cgi-bin/Tools/w3nh/w3nh.pl}) is 
0.34$\times$10$^{22}$ cm$^{-2}$ (Kalberla et al. 2005).} to the value 
0.23$\times$10$^{22}$ cm$^{-2}$.

We decided to follow different strategies in fitting the data in order to 
study in which way (if any) the source falling on a chip gap of the pn 
affects our results.
In particular, we first fitted each model (either phenomenological 
or physical), with the same set of parameters, to the MOS 1 and MOS 2 data 
sets only. The results of the best-fit procedure correspond to the entries 
labeled as $\overline{\rm {M}}$ in Table 2. We then fitted the three EPIC 
data sets with the same spectral model normalization (see the 
entries labeled as $\overline{\rm {M,p}}$). Finally, we again fitted 
the three data sets simultaneously by requiring a different model 
normalization of the pn with respect to those of the MOS (which were 
assumed to be identical). The best-fit parameters correspond to those 
labeled as $\overline{\rm{M}}$,p in Table 2.
 
In correspondence with these entries, we give in Table 2 the component 
normalizations for both the MOS (first line) and pn (second line), 
separately.

We note that all the errors in the X--ray fit parameters are quoted at the
$90\%$ confidence level. When the reduced $\chi^2_{\nu}$ is close to the 
value of 2.0 (as in the case $\overline{\rm {M,p}}$ for the 
phenomenological fit), the errors associated with the relevant quantities 
were computed by using the {\it steppar} command within XSPEC, by 
requiring that the $\chi^2_{\nu}$ increases by at least 2.71 (Avni 
1976).

As one can see from Table 2, the model parameters obtained with the 
different strategies are equivalent within the errors and, in the case of 
the phenomenological model, similar to those reported by Nucita et al. 
(2007). We also note that the pn component normalizations are 
slightly lower than those of the two MOS cameras, an effect likely due to 
the source target being located on a chip gap of the pn camera.

In each case, we estimated the observed flux in the 0.2--10 keV band and 
the equivalent width (EW) of the line at $\simeq$6.6 keV. These 
quantities are given in Table 3, following the same format as above, 
where, again, the reported errors are at 90\% confidence level.

\begin{table}
\caption[]{Estimates of the 0.2--10 keV flux and line EW.}
\vspace{-.3cm}
\scriptsize
\begin{center}
\begin{tabular}{lll}
\noalign{\smallskip}
\hline
\hline
\noalign{\smallskip}
\multicolumn{3}{c}{Phenomenological model}\\
\hline
                         &F$_{{\rm 0.2-10~keV}}$   & EW    \\
                         &(erg s$^{-1}$ cm$^{-2}$) & (keV) \\    
\hline
\multicolumn{3}{c}{}\\
$\overline{\rm{M}}$      &  $(5.4_{-0.3}^{+0.2}) \times 10^{-12}$& $0.3^{+0.2}_{-0.2}$ \\
\hline
\multicolumn{3}{c}{}\\
$\overline{\rm{M,p}}$   &  $(5.1_{-0.1}^{+0.1}) \times 10^{-12}$& $0.3^{+0.1}_{-0.1}$ \\
\hline
\multicolumn{3}{c}{}\\
$\overline{\rm{M}}$,p   &  $(5.4_{-0.3}^{+0.2}) \times 10^{-12}$& $0.3^{+0.2}_{-0.1}$ \\
\hline
\multicolumn{3}{c}{}\\
\multicolumn{3}{c}{Physical model}\\
\hline
\hline
\multicolumn{3}{c}{}\\
$\overline{\rm{M}}$      &  $(5.3_{-0.7}^{+0.4}) \times 10^{-12}$& $0.4^{+0.2}_{-0.2}$ \\
\hline
\multicolumn{3}{c}{}\\
$\overline{\rm{M,p}}$   &  $(4.8_{-1.2}^{+0.2}) \times 10^{-12}$& $0.5^{+0.3}_{-0.2}$ \\
\hline
\multicolumn{3}{c}{}\\
$\overline{\rm{M}}$,p   &  $(5.3_{-0.9}^{+0.2}) \times 10^{-12}$& $0.4^{+0.3}_{-0.2}$ \\
\hline
\hline
\noalign{\smallskip}
\end{tabular}
\end{center}
\end{table}

For completeness (and as an example), the spectra and the associated best 
fits are given in Fig. 4 for the phenomenological (left) and physical 
model (right), respectively. Here we show the fits corresponding to the 
case of different normalizations between the MOS and pn cameras 
($\overline{\rm{M}}$,p case). The red and black data points (and 
associated solid lines representing the best-fit models) refer to MOS 1 
and MOS 2, respectively. The green data points and the superimposed line 
correspond to the pn.

The results of the spectral analysis imply that the unabsorbed 0.2--10 keV 
flux of the source is $\simeq$7$\times$10$^{-12}$ erg s$^{-1}$ cm$^{-2}$. 
In the next section, we use the X--ray flux thus obtained together 
with the estimate of the source distance to get the intrinsic luminosity 
and, consequently, classify 1RXS J180431.1$-$273932.

\begin{figure*}[!t]
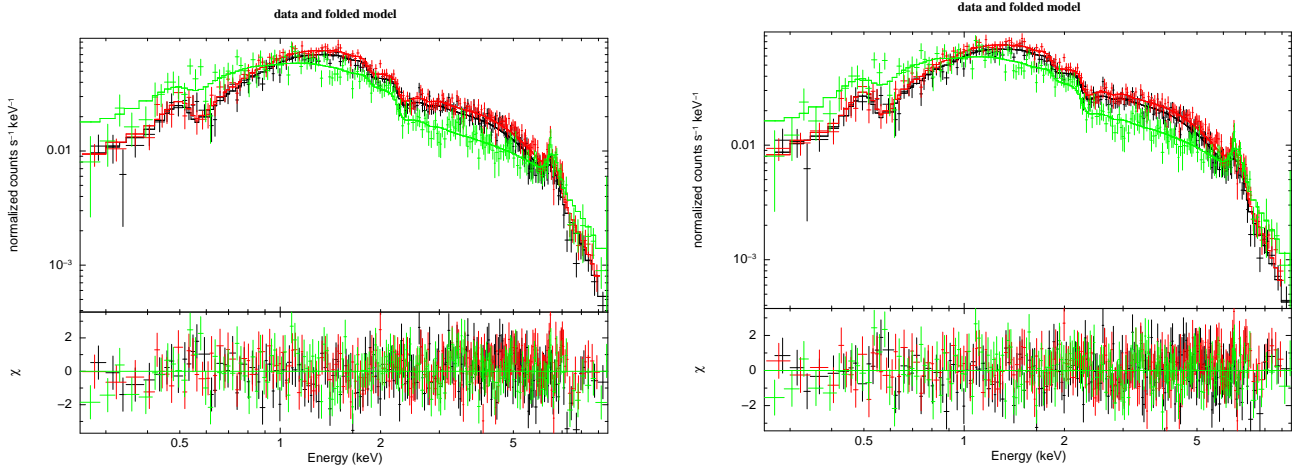

\hspace{.6cm}
\psfig{figure=19334f4l.ps,width=9cm,angle=-90}
\psfig{figure=19334f4r.ps,width=9cm,angle=-90}
\vspace{-1cm}
\caption[]{{\it XMM-Newton} X--ray spectra of 1RXS J180431.1$-$273932 and 
associated best fit lines superimposed for the phenomenological model 
(left) and the physical model (right). See text for details of 
the two models. Residuals are reported at the bottom of each panel. 
Red and black data points refer to MOS 1 and MOS 2, respectively. Green 
data points are those of the pn camera. \\
{\it (The color version of this figure is available in the on-line 
journal only)}}
\label{spectra}
\end{figure*}

The study of the X--ray light curve allowed us to confirm the existence
of the periodic signal at 494.0 s first found by Nucita et al. (2007).
We then tested the confidence level of the detected periodicity by mean of 
Monte Carlo simulations under the reasonable null hypothesis of white 
noise. The results of the analysis are shown in Fig. 5 where we 
give the Lomb-Scargle periodogram. Here, the solid, dotted, and dashed 
horizontal lines represent the 68\%, 90\%, and 99\% confidence levels 
resulted from the Monte Carlo simulations. The error associated with the 
detected period was estimated by fitting the X--ray light curve with a 
sine function and keeping the trial periods fixed. By requiring that the 
chi-square value changes of $\Delta \chi^2$ = 6.63 (see e.g. Carpano et 
al. 2007), we found the 3$\sigma$ error in the 494.0 s pulse period to be 
0.1 s. We note that the detected periodicity may be associated with the 
spin of a compact accretor, most likely a WD (see Sect. 4 for further 
details about this conclusion).

\begin{figure}
\psfig{figure=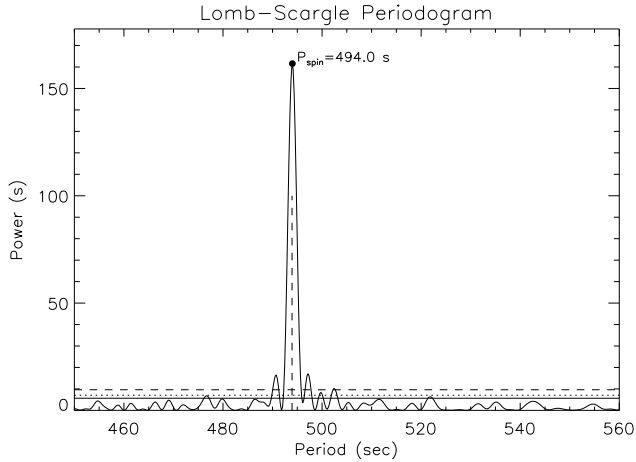,width=8.9cm,angle=0}
\vspace{-0.3cm}
\caption[]{Lomb-Scargle periodogram of the 0.3--8 keV band light 
curve of 1RXS J180431.1$-$273932 with the indication of the detected 
periodicity. The solid, dotted, and dashed 
horizontal lines represent the 68\%, 90\%, and 99\% confidence levels 
resulted from the Monte Carlo simulations (see text for details).}
\label{lomb}
\end{figure}

In Fig. 6, we give the total (MOS+pn) 0.3--8 keV band light curve 
folded at the detected period and with 20 bins. The zero phase is 
associated with the beginning of the {\it XMM-Newton} observation.

\begin{figure}
\psfig{figure=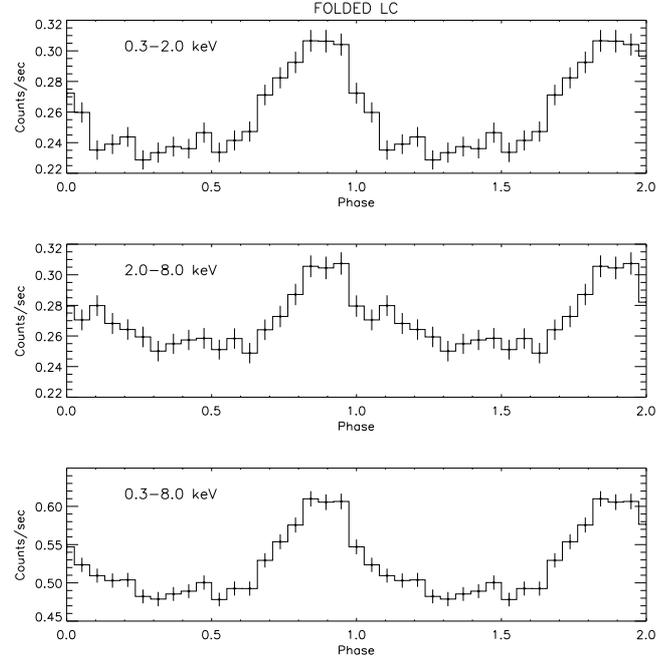,width=8.9cm,angle=0}
\vspace{-0.3cm}
\caption[]{Total (MOS+pn) 0.3--8 keV band light curve folded at 
494.0 s using 20 bins (see text for details).}
\label{folded}
\end{figure}

The light curve was then binned at twice the detected periodicity (see 
Fig. 7) to search for variability in the X--ray signal. By fitting the 
light curve with a simple linear function, we found a clear trend in the 
data (as already noted in Nucita et al. 2007) corresponding to a decrease 
in the count rate of $\sim 10^{-4}$ counts s$^{-1}$ hr$^{-1}$, and a 
variability on the timescale of hours.

\begin{figure}[!th]
\psfig{figure=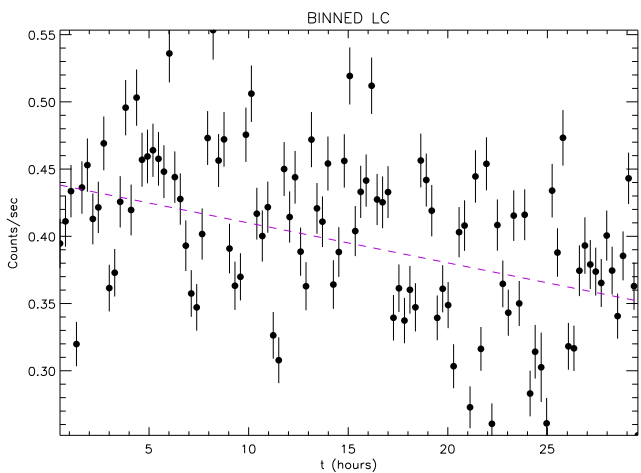,width=8.9cm,angle=0}
\vspace{-0.3cm}
\caption[]{MOS+pn 0.3--8 keV light curve binned at twice the 494.0 s 
periodicity (see text for details), together with the best linear fit 
superimposed (dashed line).}
\label{binned}
\end{figure}

\section{Discussion}

Using optical imaging and spectroscopy, we investigated two sources in the
X--ray error circle of 1RXS J180431.1$-$273932 and its proximity.
We found that the one lying within the circle is a magnetic CV, while the
other, slightly outside the error box, is a red giant, as correctly 
inferred by Nucita et al. (2007).

This latter finding led those authors to propose that this X--ray source 
could be a new member of the small class of SyXBs; however, the presence 
of a magnetic CV within the error circle and the positional displacement 
between the red giant and the X--ray position led us to conclude that the 
actual X--ray emitter is the CV.

Magnetic CVs are known to be X--ray sources and, 
thanks to the {\it INTEGRAL} and {\it Swift} missions many of them 
were discovered or detected at energies above 20 keV (Barlow et al. 2006; 
Brunschweiger et al. 2009; Landi et al. 2009; Scaringi et al. 2010).
The spin period range of the WD hosted in these systems, in 
particular the ones in IPs (see e.g. Butters et al. 2011) encompasses the 
494 s X--ray periodicity detected by Nucita et al. (2007) from 1RXS 
J180431.1$-$273932, which can thus be interpreted as such. 
The iron emission line around 6.6 keV with an EW of several 
hundreds of eV is a typical X--ray spectral characteristic of magnetic 
CVs (de Martino et al. 2008). In the same vein, we see that the 0.2--10 
keV band softness-ratio (as defined in Ramsay \& Cropper 2004) computed 
using the unabsorbed fluxes of the black-body ($\sim$5.5$\times$10$^{-13}$ 
erg cm$^{-2}$ s$^{-1}$) and thermal bremsstrahlung 
($\sim$6.5$\times$10$^{-12}$ erg cm$^{-2}$ s$^{-1}$) components is in the 
present case $\sim$0.02, a value typical of IP systems (Evans \& 
Hellier 2007). To all this, one can add that the bremsstrahlung 
temperature and the 0.2--10 keV luminosity of the source (which is 
1.7$\times$10$^{32}$ erg s$^{-1}$ assuming the distance determined 
in the previous section) are comparable with those typical of these sources 
(as can be determined from, e.g., Landi et al. 2009). All this supports 
the identification of this X--ray source as a magnetic CV.

We stress that it is known (see e.g. Masetti et al. 2006a, 2007) that the 
optical spectra of red giant companions in SyXBs do not generally show any 
peculiarity such as emission lines (the only notable exception being GX 
1+4: Chakrabarty \& Roche 1997), so on this basis the object 2MASS 
J18043013$-$2739340 cannot be ruled out as the optical counterpart of 1RXS 
J180431.1$-$273932. However, since (i) it lies nominally outside 
the X--ray error circle of this high-energy source, and (ii) a magnetic CV 
is actually found inside this circle, we can confidently state that this 
X--ray source is not a SyXB, but rather a magnetic CV, likely of IP type, 
and that the red giant is just a fore-/background object.

To conclude, we can use some of the results from the physical description 
of the X--ray spectrum to infer parameters relative to the X--ray 
emitter. In particular, using the bremsstrahlung component temperature 
($kT_{\rm br} \sim$ 40 keV) and Eq. (3) of Middleton et al. (2012), we 
obtain a mass $M_{\rm WD}$ = 0.8$^{+0.4}_{-0.3}$ $M_\odot$ for the 
accreting WD hosted in this system (the quoted errors are at 90\% 
confidence level). This, considering the X--ray luminosity of the source 
and assuming a radius $R_{\rm WD} \sim$ 6700 km for a WD with mass 0.8 
$M_\odot$ (Nauenberg 1972), implies an 
average mass accretion-rate $\dot{m} \sim$ 1.6$\times$10$^{-11}$ 
$M_\odot$ yr$^{-1}$ for the source. Likewise, from the best-fit value of 
the black-body normalization we determine a radius $r_{\rm bb} \sim$ 1 km
for the area of this emission component. This value is quite modest when 
compared with the size of the WD surface; however, it is not atypical of
magnetic CVs (see e.g. Anzolin et al. 2008).

Again following Anzolin et al. (2008), and accepting the presence around 
the WD of an accretion disk truncated at the magnetospheric radius $r_{\rm 
mag}$, we can assume that this quantity is comparable in size with the 
corotation radius $r_{\rm co}$, which is the radius at which the magnetic 
field of the WD rotates with the same Keplerian frequency of the inner 
edge of the accretion disk. According to this hypothesis, we can determine 
the magnetic moment of the WD hosted in 1RXS J180431.1$-$273932 to be $\mu 
\sim$ 3.7$\times$10$^{32}$ G cm$^3$.

\section{Conclusions}

Our multiwavelength optical/X--ray study of 1RXS J180431.1$-$273932
has allowed us to identify its actual counterpart and pinpoint its
real nature. We have found that this object is a magnetic CV, most likely 
of IP type; the SyXB hypothesis, put forward by Nucita et al. (2007), is 
thus ruled out. This misidentification was likely induced by the presence 
of a red giant lying along the line of sight and just outside the 
border of the X--ray error box. We also exclude any connection of 
the X--ray source with other, fainter optical objects within its 
positional uncertainty obtained from {\it XMM-Newton} data.

We have confirmed the X--ray periodicity of 494 s first detected by Nucita 
et al. (2007) and interpreted it as the spin period of the accreting WD 
hosted in this system. We could also successfully model the X--ray 
spectrum of 1RXS J180431.1$-$273932 using a bremsstrahlung plus black-body 
emission model, as typically found in magnetic CVs. We encourage the 
acquisition of follow-up observations in the optical and X--rays to 
determine the orbital period and the main physical characteristics of this 
CV, to also confirm the IP nature proposed here.

This research moreover stresses that it is of paramount importance to have 
very precise (smaller than a few arcsec) X--ray localizations, especially 
in cases of crowded fields and particularly for objects concentrated 
towards the Galactic bulge, in order to determine the optical counterpart.

\begin{acknowledgements}

We thank Aldo Fiorenzano for the Service Mode optical observations 
acquired at TNG and presented in this paper, and Domitilla de Martino for
suggestions. AAN is grateful to Stefania Carpano for interesting 
discussions.
We also thank the anonymous referee for useful remarks that helped us 
to improve the quality of this paper. 
This research has made use of the NASA Astrophysics Data System Abstract 
Service and the NASA/IPAC Infrared Science Archive, which are operated by 
the Jet Propulsion Laboratory, California Institute of Technology, under 
contract with the National Aeronautics and Space Administration. 
This publication made use of data products from the Two Micron All 
Sky Survey (2MASS), which is a joint project of the University of 
Massachusetts and the Infrared Processing and Analysis Center/California 
Institute of Technology, funded by the National Aeronautics and Space 
Administration and the National Science Foundation.
This paper is also based on observations from {\it XMM-Newton}, an ESA 
science mission with instruments and contributions directly funded by ESA 
Member States and NASA. NM acknowledges financial contribution from the 
ASI-INAF agreement No. I/009/10/0. 
\end{acknowledgements}

\end{document}